\documentclass[epj]{svjour}
\pdfoutput=1
\usepackage{amsfonts,amsmath,amssymb}
\usepackage{graphicx}
\usepackage{color}
\usepackage{lipsum}
\usepackage[plainpages=false, colorlinks=true, anchorcolor=blue, linkcolor=blue, citecolor=blue, bookmark
s=false]{hyperref}

\begin{document}

\newcommand{\apjl}{Astrophys. J. Lett.}
\newcommand{\apjs}{Astrophys. J. Suppl. Ser.}
\newcommand{\aap}{Astron. \& Astrophys.}
\newcommand{\aj}{Astron. J.}
\newcommand{\apj}{Astrophys. J.}
\newcommand{\araa}{Ann. Rev. Astron. Astrophys. } 
\newcommand{\mnras}{Mon. Not. R. Astron. Soc.}
\newcommand{\solphys}{Solar Phys.}
\newcommand{\jcap}{JCAP}
\newcommand{\pasj}{PASJ}
\newcommand{\pasa}{Pub. Astro. Soc. Aust.}
\newcommand{\apss}{Astrophysics \& Space Science}
\newcommand{\aaps}{Astron. Astrophys. Suppl. Ser.}
\newcommand{\Dtwoo}{$\mathrm{D_2O}$ }
\newcommand{\prd}{Phys. Rev D. }

\title{Generalized Lomb-Scargle analysis of   $\rm{^{90}Sr/^{90}Y}$ decay rate measurements from the Physikalisch-Technische Bundesanstalt}

\author{Tejas  P.\inst{1}  \and   Shantanu  Desai\inst{1} }

\institute{$^{1}$ Department of Physics, IIT Hyderabad, Kandi, Telangana 502285 India}

\authorrunning{Tejas.P and S. Desai} 
\titlerunning{Generalized L-S analysis of $\rm{^{90}Sr/^{90}Y}$ decay measurements  }

\abstract{
We apply the generalized Lomb-Scargle (LS) periodogram  to  independently confirm   the  claim by  Sturrock et al~\cite{Sturrock16} of  an oscillation at a frequency of 11/year  in the decay rates of $\rm{^{90}Sr/^{90}Y}$ from measurements at the  Physikalisch Technische Bundesanstalt (PTB),  which however has been disputed by Kossert and Nahle~\cite{Kossert}. For this analysis, we made two different {\it ansatze}  for the errors. For each peak in the LS periodogram, we evaluate the statistical significance using non-parametric bootstrap resampling. \textcolor{black}{We find using  both of these error models evidence for ~11/year periodicity in the $\rm{^{90}Sr/^{90}Y}$ data for two of the three samples, but at a lower significance than that claimed  by Sturrock et al~\cite{Sturrock16}.}}

                                                     \PACS{
      {23.40.-s}{Beta Decay} \and
      {02.50.-r}{Statistics, 02.50.-r}     
     }
\maketitle

\section{Introduction}

In the past two decades, there have been a number of works starting with Falkenberg~\cite{Falkenberg} pointing out that the beta decay rates are variable and depend on various environmental parameters. Some of the environmental influences proposed for this variability include  solar rotation, other ancillary dynamics in the inner solar core~\cite{Sturrock12,Sturrocksolar}, solar flares~\cite{Jenkins}, Earth-Sun distance~\cite{Jenkins2}, lunar influence etc~\cite{Parkhomov}.
However,  these results have been disputed by other authors (eg.~\cite{Pomme,bellotti2012,bellotti2,bellotti5})  and no common consensus has emerged. A summary of some of these claims as well as their rebuttals  are reviewed in Refs.~\cite{Sturrock16,Kossert,Pomme}.

In this work we concentrate on settling the contentious claim of one such result regarding the decay rates of $\rm{^{90}Sr/^{90}Y}$ from one specific experiment, between two groups of authors. Parkhomov~\cite{Parkhomov} and Sturrock et al~\cite{Sturrock12} found evidence for annual and monthly oscillations in the decay rates of $\rm{^{90}Sr/^{90}Y}$ measured at Institute for Time Nature Explorations, Moscow State University. Furthermore,   Sturrock et al~\cite{Sturrock12} also found correlations between these decay rates and r-mode oscillations inside the Sun.

These results were contested by Kossert and Nahle~\cite{Kossert} (hereafter, KN15). They showed using long-term measurements of the decay rates with a custom-built liquid scintillator at  the Physikalisch-Technische Bundesanstalt (PTB), 
that there is no evidence for any periodic modulations in the decay rates of $\rm{^{90}Sr/^{90}Y}$. The results of KN15 were in turn rebuked by Sturrock et al~\cite{Sturrock16}  (hereafter, S16), who reanalyzed the same PTB data from KN15 and found evidence for statistically  significant peaks at 11/year.
S16 further argued that this oscillation  frequency is indicative of a solar influence.

S16 used a likelihood procedure~\cite{Sturrock05} analogous to the Lomb-Scargle periodogram to analyze the data and found peaks at the same location as KN15. However, the $p$-values they obtained (of the peaks been a random fluctuation) were much smaller than in KN15, implying an enhanced statistical significance for the peaks.
One criticism of the KN15 paper by S16 was that KN15 incorrectly calculated the significance of each peak as $\exp(-\sqrt{S})$, instead of $\exp(-S)$, where $S$ is the LS power. The significance of the peaks was also independently validated  by S16 using a shuffle test~\cite{Bahcall91}.

Here, we focus on adjudicating the above  conflict between  KN15 and S16 regarding  the oscillations in the decay rates of  $\rm{^{90}Sr/^{90}Y}$ at PTB, which remains unresolved,  using an independent analysis and with a slight variant of their analysis.
For this purpose, we use a modified version of the Lomb-Scargle periodogram called ``Generalized Lomb-scargle periodogram" (or floating-mean periodogram) to analyze the same dataset and evaluate the significance using bootstrap resampling. The same procedure was previously used in particle physics to assess the significance
of periodicity in solar neutrino flux measured in Super-Kamiokande and SNO experiments~\cite{Desai16}. However, this generalized  periodogram is  routinely used throughout  astronomy (for example, see ~\cite{Anderson}).

The outline of this paper is follows. The generalized Lomb-Scargle periodogram is introduced in Sect.~\ref{sec:ls}. Our analysis of the PTB is described in Sect.~\ref{sec:analysis}. A comparison of our results with those of Sturrock et al can be found in Sect.~\ref{sec:scomp}. We then address the question of whether the observed data is purely stochastic in Sect.~\ref{sec:stochastic}.
We conclude in Sect.~\ref{sec:conclusions}.

\section{Generalized Lomb-Scargle Periodogram}
\label{sec:ls}
The Lomb-Scargle (hereafter, LS)~\cite{Lomb,Scargle} (see Ref.~\cite{Vanderplas} for a recent review) periodogram is a widely used technique in astronomy and particle physics to look for  periodicities in unevenly sampled datasets, and has been applied to a large number of astrophysical datasets. Here,  for our analysis,  we shall apply a slight variant of the normal LS periodogram. We first provide a bare-bones introduction to the normal LS periodogram and then briefly outline  the modification proposed by Zechmeister and Kurster~\cite{Kurster}, which is known in the literature as the generalized LS periodogram~\cite{Bretthorst,Kurster} or the floating mean periodogram~\cite{Cumming,Vanderplas15,Vanderplas} or the Date-Compensated Discrete Fourier Transform~\cite{Ferrazmello}. More details are outlined in Refs.~\cite{Vanderplas,astroml} and references therein.

The goal of the LS periodogram  is to determine the angular frequency ($\omega$) of a  periodic signal in a time-series dataset  $y(t)$
given by:
\begin{equation}
y(t)=a\cos(\omega t)+ b \sin(\omega t).
\label{eq:yt}
\end{equation}
It can be obtained as an analytic solution,  while solving the problem of   fitting for a sinusoidal function by $\chi^2$ minimization, 
 and hence is a special case of the maximum likelihood technique for symmetric errors~\cite{Ranucci}. 
The LS periodogram calculates the power as a function of frequency, from which one needs to infer the presence of a sinusoidal signal.

One premise in calculating the  LS periodogram~\cite{Lomb,Scargle} is that the data are pre-centered around the mean 
value of the signal. This pre-centering is done using the sample mean, which is computed from the existing data. One {\em ansatz} implicitly made is that 
this is  a good estimate for the mean value of the fitted function. This assumption breaks down  if the data does not uniformly sample all the phases, or if the dataset is small and does not extend over the full duration of the sample. Such errors in estimating the mean can cause aliasing problems~\cite{Kurster}.
Therefore, to circumvent these issues, the LS periodogram 
was generalized to add an arbitrary offset to the mean values~\cite{Kurster} as follows:
\begin{equation}
y(t)=y_0 (f) +  a\cos(\omega t)+ b \sin(\omega t),
\end{equation}
\noindent where $y_0(f)$ is an offset term  added to the sinusoidal model at each frequency
We refer to this modification  as the ``generalized'' LS periodogram in the remainder of this  work.   But as mentioned earlier, this modification is also referred elsewhere in literature as  the floating-mean periodogram.
The resulting equations for the generalized LS power can be
found in  Eq. 20 in  Ref.~\cite{Kurster}. 
It has been shown that the generalized LS periodogram is more sensitive than the normal one in detecting periodicities, in case the data sampling overestimates the mean~\cite{Vanderplas,Kurster,proc}. In this work, we shall use the generalized LS periodogram for all the analyses.

If the observed data show any sinusoidal modulations at a given frequency, one would expect a peak in the LS periodogram at that frequency. To assess the significance of such a  peak, we use the bootstrap method, in which for the same temporal coordinates as the data, we draw points  randomly with replacements from the observed values and recompute the periodograms. Such a  non-parametric bootstrap resampling procedure   can reproduce any empirical distribution along with extreme-value methods to account for the tails~\cite{Suveges}. To assess the significance of any peak, we  shall compute the significance using 1000 bootstrap resamples of the data.  

\section{Analysis}
\label{sec:analysis}

\subsection{Dataset}
The PTB dataset consists of three samples of $\rm{^{90}Sr/^{90}Y}$ denoted as S2, S3, and S4. This is supplemented by a blank sample  (S1) for monitoring the background effects. The radioactivity estimates have been made using the Triple-To-Double coincidence ratio method~\cite{Kossert}. The beta decay rates are parameterized by the normalized activity rates as shown in Figures 4, 5, and 6 of KN15. The normalization takes into account the triple coincidence rate and counting efficiency.
More details of the sample preparation and the PTB measurements can be found in KN15. 

\subsection{Power spectrum analysis}
\begin{center}
\begin{figure*}
\includegraphics[scale=0.5]{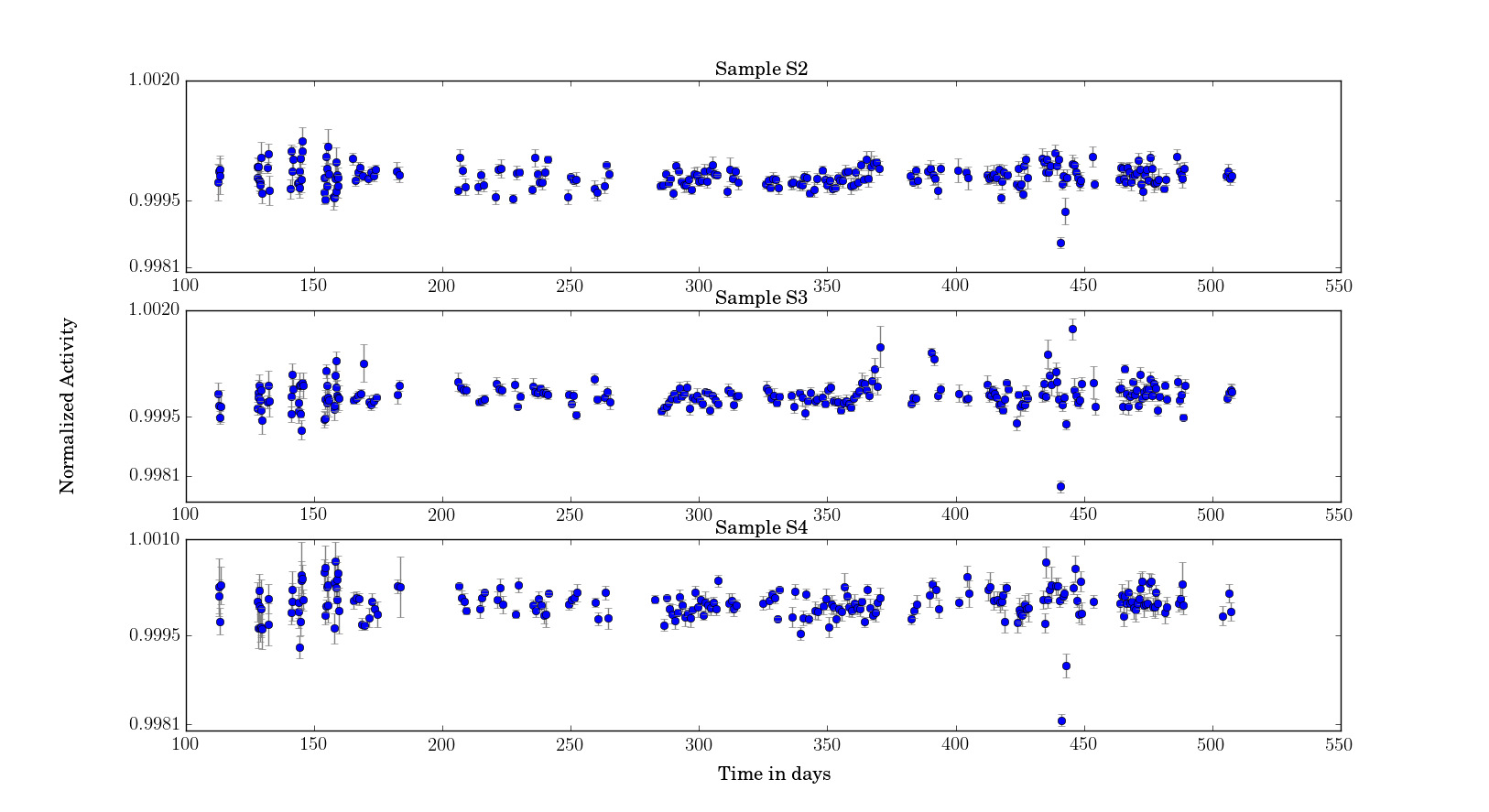}
\caption{Plot showing representative activity values and representative time instances of all the 240 bins for each of the three samples. The plot also shows the error in each bin assuming standard error of the mean.}
\label{timeseries}
\end{figure*}
\end{center}

We have used the generalized Lomb-Scargle periodogram  to detect a possible periodicity  in the unevenly sampled activity data. The activity data from PTB contain many time periods without any data. The data \textcolor{black}{were} organized into bins and clustered. Contiguous data points were grouped into a single bin. All the data points in a bin were clustered, that is, the data points were replaced with a single value representative of all the points in that bin. After clustering the data, we obtained 240 time bins. Since there were no error bars provided per data point,  we computed the periodogram  by positing two different error models: For the first analysis we assumed that the error in each bin is given by the  standard error of the mean, which is similar to the analysis done in S16. A time series representation of the data for all the three samples with this error model is shown in Fig.~\ref{timeseries}.
We also redid this analysis assuming an error of 0.03\% per data point. This is the average error estimated by KN15 (Table 1), from a quadrature sum of the different sources of systematic errors. We note however that it is  not explicitly stated in KN15 as to whether this particular error budget has  been  used for their  periodograms for the three samples.  We also couldn't find any information on the bin size used for the periodogram  analysis carried out in KN15. On the other hand, S16 grouped the data into 50 bins of equal occupancy.

We computed the generalized LS periodogram using the {\tt lomb\_scargle} routine from the  {\tt astroML}~\cite{astroml} Python library. From the LS power at different frequencies,  we need to estimate the significance at that frequency. 
The minimum and maximum frequencies have to be carefully chosen. The spacing between the frequencies has to be chosen so as to not miss any peaks. The choice of minimum frequency is straight-forward, $f_{min}=1/T$, where $T$ is the total time spanned by the set of observations. For all practical purposes, the minimum frequency is chosen to be zero. The maximum frequency is suggested as $(1/T_{med})$, where $T_{med}$ is the median of the difference of the representative time instances of each bin \cite{Vanderplas}. The median time between consecutive bins after choosing a time bin to be the contiguous data  is equal to approximately 1.006 days. The reason for the slight variation in this value for different samples is due to a very small variation in the bin sizes and time instances corresponding to the different data points.

Since our main goal  is to resolve the conflicting claims in two papers, we restrict ourselves to a maximum frequency of $20 \ yr^{-1}$,  as in KN15 and S16 instead of the maximum permissible frequency. The spacing between the successive frequencies is chosen to be $1/(5T)$ and is equal to 5.87 x 10$^{-9}$ Hz. Note that  there is a slight variation of this value for different samples due to a very small variation in the bin sizes and time instances corresponding to these data points. (For more details on these recommendations, see Vanderplas~\cite{Vanderplas} and references therein).

So, the recommended minimum and maximum values of angular frequency $\omega$, which we use are: $\omega_{min}=(2\pi /T)$ and $\omega_{max}=2\pi f_{max}$, where $f_{max}= 20 \ yr^{-1}$. We choose the value of $\omega_{min}$ very close to zero (1 x 10$^{-9}$ rad/sec) here, instead of zero exactly, because the LS routine does not return a valid value for zero frequency.   The total number of frequencies at which the power is computed is equal to $(\omega_{max} - \omega_{min})/(2\pi /5T)$. The total number of frequencies at which power is computed corresponds to 108, 108 and 107 for samples S2, S3 and S4 respectively.

We now report results from both these analyses.

\subsubsection{Analysis assuming 0.03\% error per data point}
\label{sec:anal1}
For the first error model, the representative activity value of each bin was calculated from  the weighted mean of the observed data, wherein each data point is  weighted by the  inverse square of the error. The representative activity value ($a_{b}$) and time instance ($t_b$) of a given bin are computed as:
\begin{eqnarray}
a_b &=& \frac{\sum_{i}d_i/(e_i)^{2}}{\sum_{i}1/(e_i)^{2}} \\
e_i &=&  \frac{0.03}{100}d_i \\
t_b &=& \frac{1}{N}\sum_{i}t_i \\
\end{eqnarray}
\noindent where $N$ is the number of data points in the bin;  $e_i$ is the error per data point; and  $t_i$ is the time instance corresponding to data point $d_i$. 

The average activity error in each time bin ($\sigma_i$) is computed by propagating the uncertainty in each data point:
\begin{equation}
\sigma_i^{2}=\frac{1}{\sum_{i}\left( \frac{1}{e_i} \right)^{2}}. 
\end{equation}

Using the above error budget, we then construct the LS periodogram for each of the three samples S2, S3, and S4. For each of these samples, we show the LS power  and a horizontal line representing the False-Alarm Probability (FAP) of the most significant peak using 1000 bootstrap resamples. From the FAP, one can obtain an assessment of the statistical significance of any peak in the periodogram. For a peak to be  statistically significant indicative  of any oscillations, FAP should be as small as possible.

\begin{figure}
\includegraphics[width=8cm]{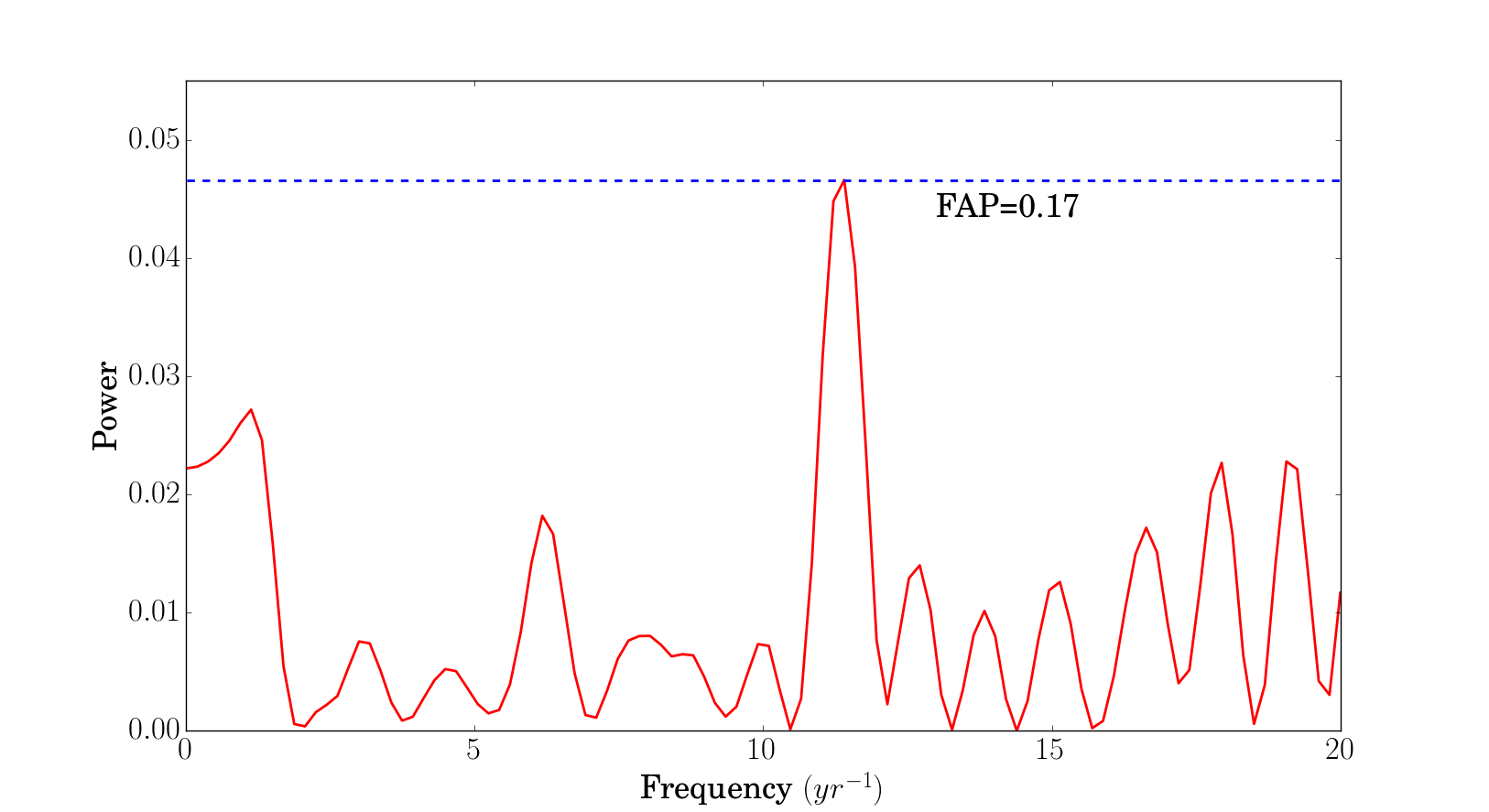}
\caption{Power spectrum of PTB Sample 2, assuming an error of 0.03\% per data point. Note that the periodograms have been normalized according to Ref.~\cite{astroml}. To recover the LS powers in  KN15~\cite{Kossert} and Sturrock~\cite{Sturrock16}, one needs to multiply by $(N-1)/2$. 
The dotted horizontal line corresponds to a false alarm probability (FAP) of a random fluctuation  equal to 17.2\% and represents the FAP of the  largest  peak in the LS periodogram. In this case, this peak is at about 11.4 $yr^{-1}$. However, the FAP at this peak is consistent with it been a noise fluctuation.}
\label{fig:S2e1}
\end{figure}

\begin{figure}
\includegraphics[width=8cm]{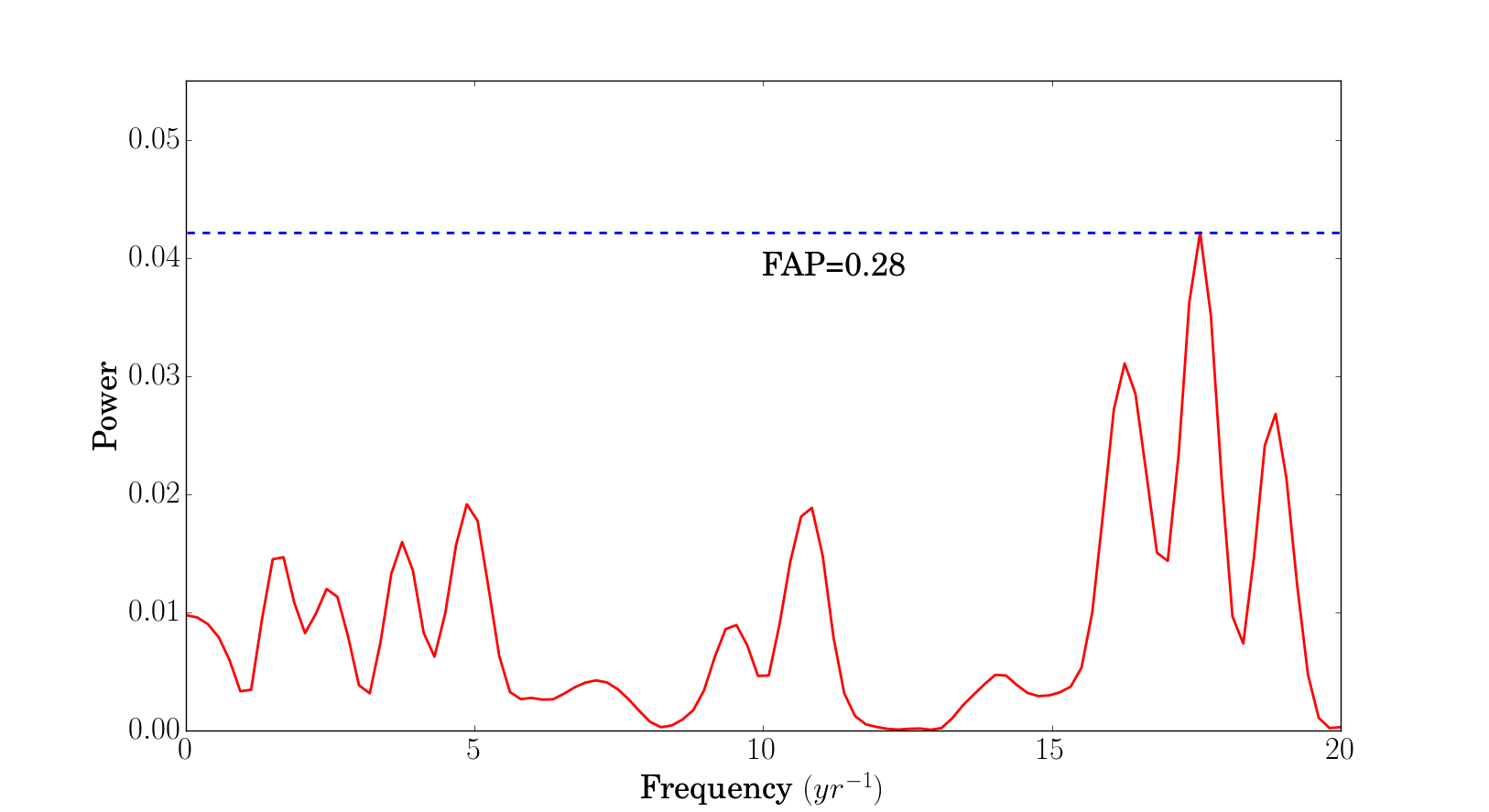}
\caption{Power spectrum of PTB Sample 3, assuming an error of 0.03\% per data point. See Fig.~\ref{fig:S2e1} for more details about the labels. The dotted horizontal line corresponds to a FAP 28.2\% and is not significant.}
\label{fig:S3e1}
\end{figure}

\begin{figure}
\includegraphics[width=8cm]{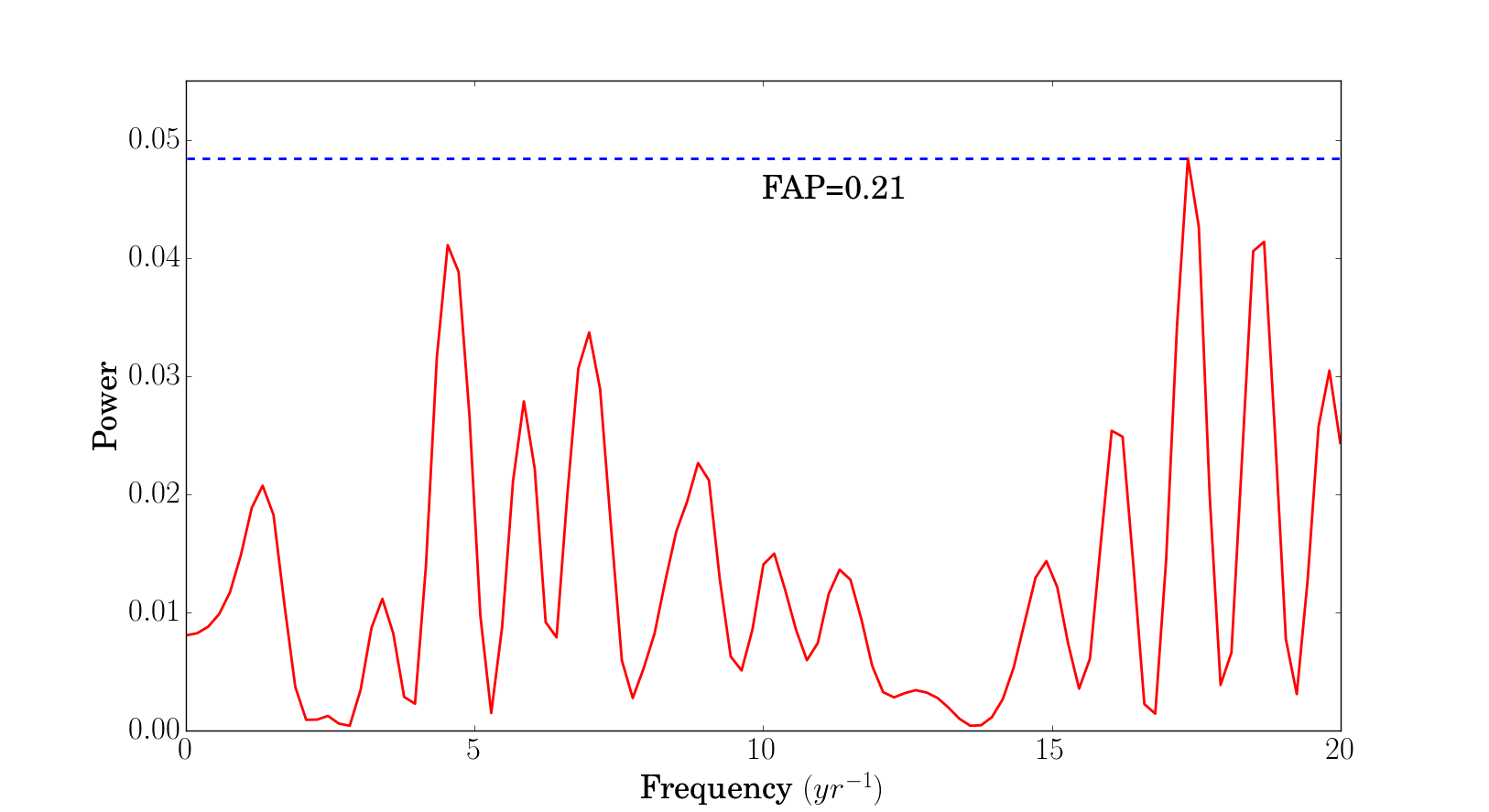}
\caption{Power spectrum of PTB Sample 4, assuming an error of 0.03\% per data point. See Fig.~\ref{fig:S2e1} for more details about the labels. The dotted horizontal line corresponds to a FAP of 20.5\% and is consistent with noise. Here, there is {\bfseries{no}} observed peak at 11.5 $yr^{-1}$.}
\label{fig:S4e1}
\end{figure}

Figs.~\ref{fig:S2e1},~\ref{fig:S3e1},~\ref{fig:S4e1} show the corresponding LS periodogram (power vs frequency in units of  $\rm{yr^{-1}}$) with these assumptions for samples S2, S3, and S4 respectively. A tabular summary of these results can be found in Table~\ref{deltabic}.

This normalization of the LS power (which follows the convention  originally proposed by Lomb~\cite{Lomb}), differs from that used in  KN15 and S16, (which follows Scargle's convention~\cite{Scargle}) by a factor of $(N-1)/2$ for $N$ data points.  With this assumption, the values for the LS power fall between 0 and 1. A tabular summary of these results can be found in Tab.~\ref{deltabic}.

 Despite using different  bin sizes, the locations of the peak frequencies in all the periodograms  with this error model is same as in KN15 and S16. For S2, the periodogram is peaked at about 11.4 /year (with FAP of about 17.2\%). S3 and S4 show peaks at 17 per year with FAPs of 28.2\% and 20.5\% respectively. Therefore, the significance of all these peaks (based on the FAP)  is marginal and cannot be construed as statistically significant evidence for oscillations at any frequency. If there is any influence from the solar interior on the beta decay rates, then all the three samples should show statistically significant peaks around 11 per year, which we do not find. The FAPs we obtain are much higher than S16 and are consistent with noise. 

\subsubsection{Analysis assuming standard error of mean}
We now re-calculate  the periodogram by positing that  the error in each bin is the  standard error of the mean, which
is similar to the analysis done in S16. However in S16, 50 bins were chosen in such a way that the number of data points in each bin were the same, whereas for our analysis, the bins represent contiguous periods of data.  
In this case, the representative activity  value ($a_b$) as well as the central time in each bin ($t_b$) were taken to be  the mean of the activity values in a bin:
\begin{eqnarray}
a_b &=& \frac{1}{N}\sum_{i}d_i  \\
t_b  &=& \frac{1}{N}\sum_{i}t_i
\end{eqnarray}

We note that $t_b$ is calculated in the same way as in our previous analysis. The error in each bin, which in this case is the  standard error $SE$ is computed as follows:

\begin{eqnarray}
SE &=&  \frac{\sigma}{\sqrt{n}} \\
\sigma &=& \sqrt{\frac{1}{N}\sum_{i}\left(d_i - \mu \right)^{2}}
\end{eqnarray}
where $\sigma$ is the standard deviation of the data points in a given bin, $\mu$ is the mean of the data points in a given bin and $N$ is the total number of data points in a given bin. 

\begin{figure}
\includegraphics[width=8cm]{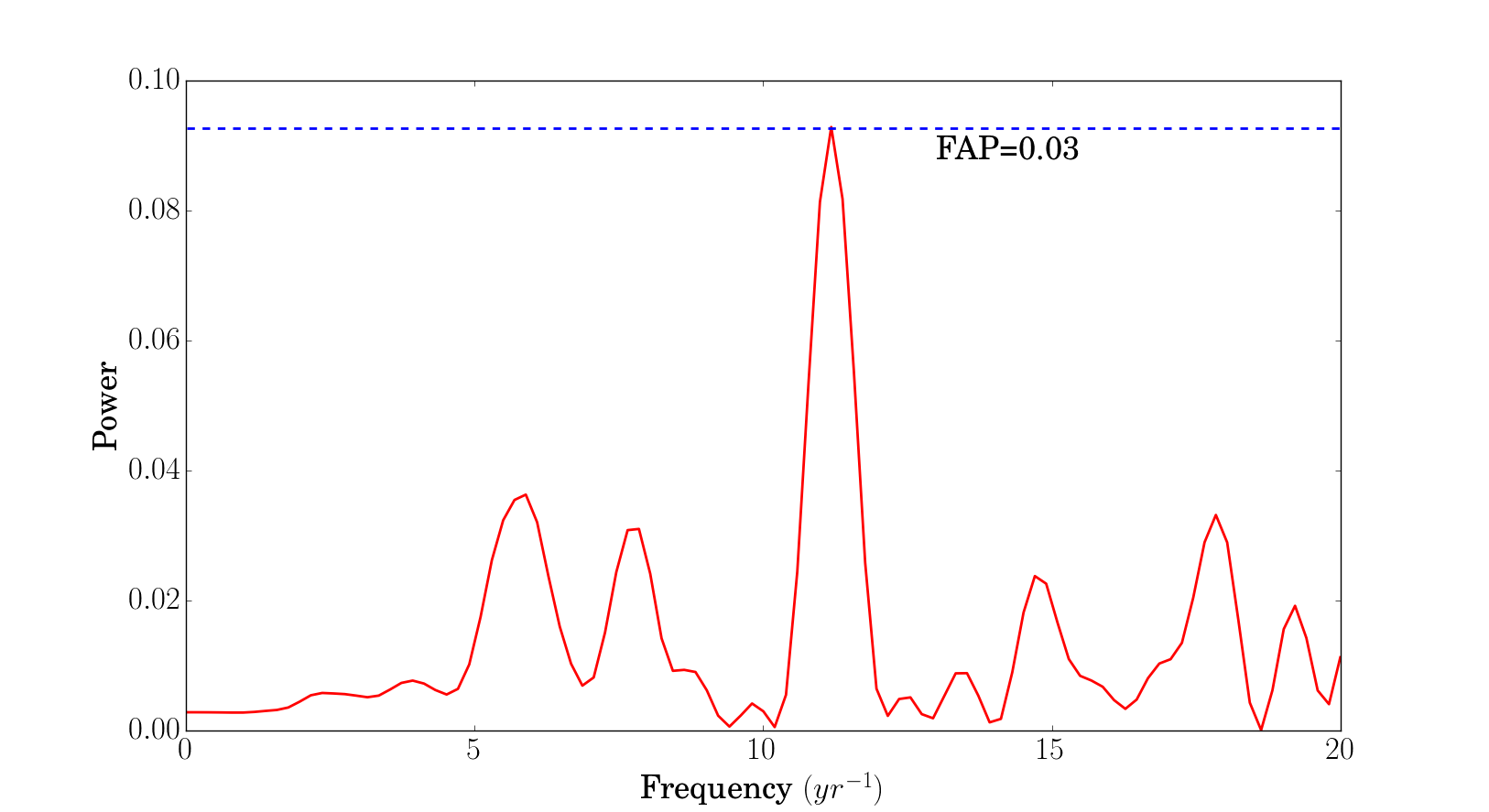}
\caption{Power spectrum of PTB Sample 2, assuming standard error of mean. See Fig.~\ref{fig:S2e1} for more details about the labels. The dotted horizontal line corresponds to a FAP of 2.6\% and corresponds to a significance of 1.94$\sigma$.}
\label{fig:S2e2}
\end{figure}

\begin{figure}
\includegraphics[width=8cm]{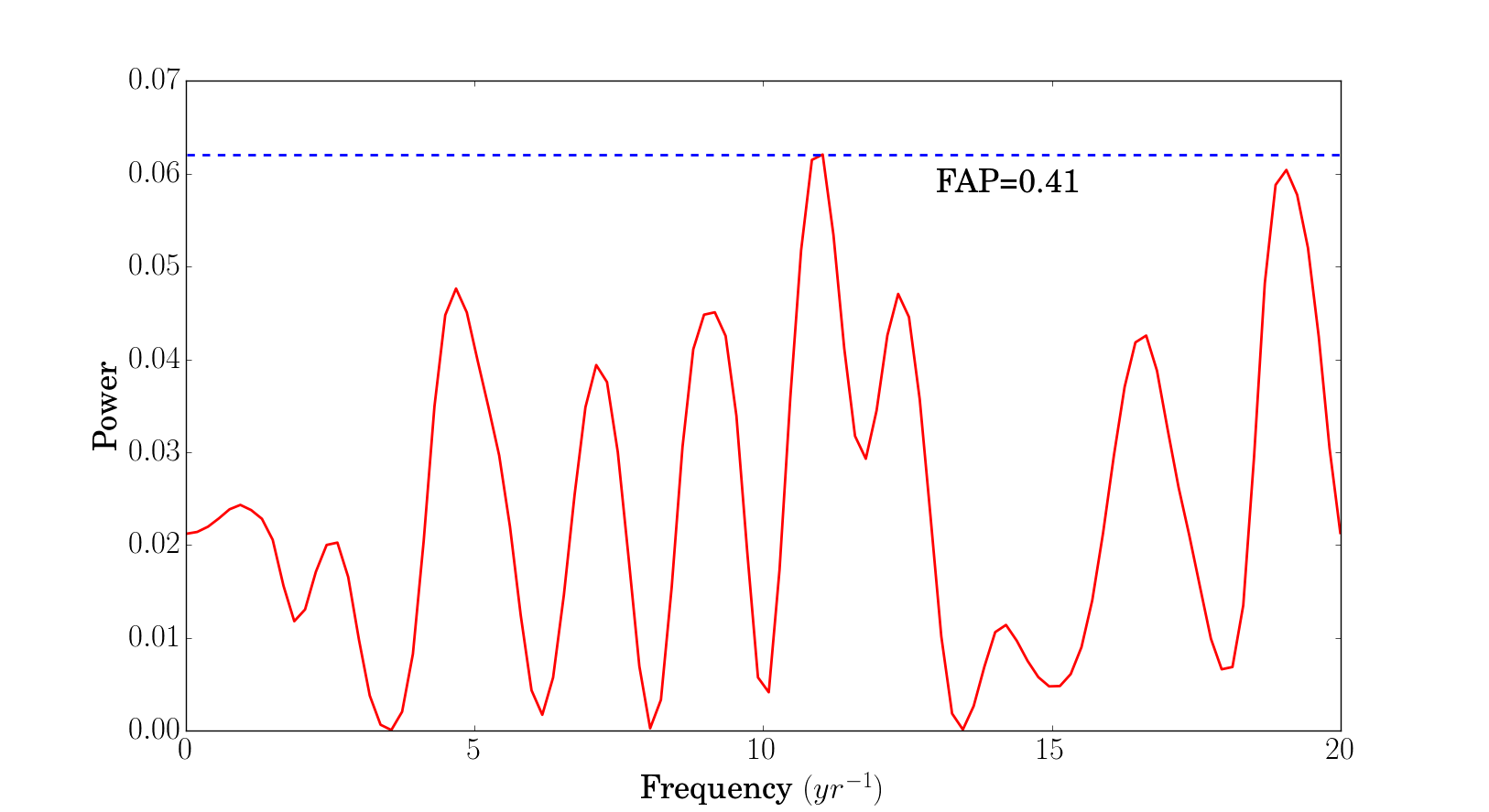}
\caption{Power spectrum of PTB Sample 3, assuming standard error of mean. See Fig.~\ref{fig:S2e1} for more details about the labels. The dotted horizontal line corresponds to a FAP of 41.4\% and is consistent with noise.}
\label{fig:S3e2}
\end{figure}

\begin{figure}
\includegraphics[width=8cm]{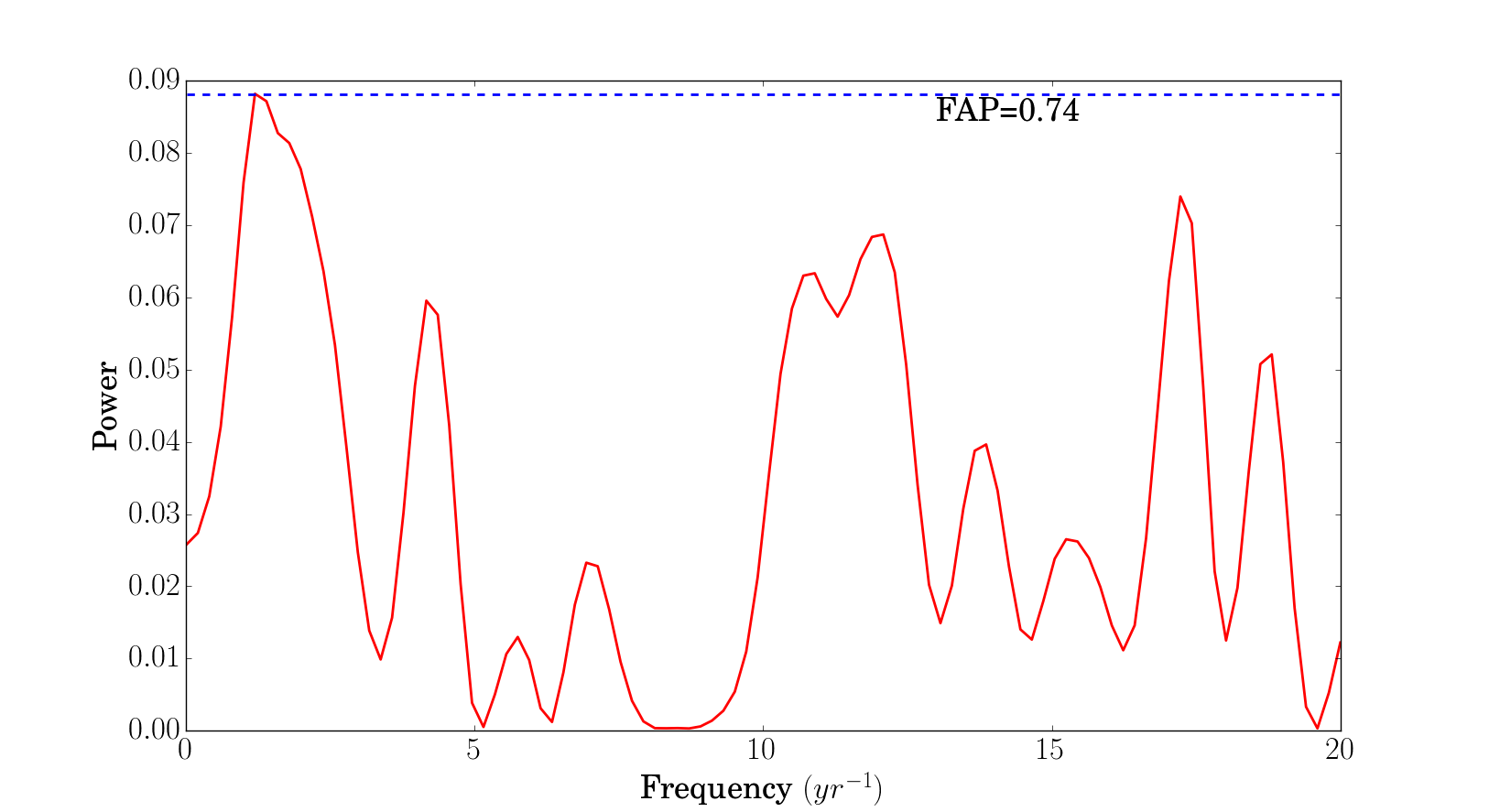}
\caption{Power spectrum of PTB Sample 4, assuming standard error of mean. See Fig.~\ref{fig:S2e1} for more details about the labels. The dotted horizontal line corresponds to a FAP of 74.1\% and is consistent with noise. }
\label{fig:S4e2}
\end{figure}

Using this error budget for each data point, we then construct the LS periodograms in the same way as before. These periodograms can be found in Figs.~\ref{fig:S2e2},~\ref{fig:S3e2} and ~\ref{fig:S4e2} respectively. A tabular summary of the results with this model for the error budget can be found in Table~\ref{deltabic}. This time, we find that both S2 and S3 show  peaks  at a  frequency of approximately 11/year. S4 shows a peak at about 1.3  per year. Therefore, the location of the peak frequencies in samples S3 and S4 is different than our previous analysis in Sect.~\ref{sec:anal1} as well as with the results from KN15 and S16. However, even in this case  none of the peaks are statistically significant. The FAP of the peaks for S2, S3, and S4 are 2.6\%, 41.4\% and 74.1\%. The lowest FAP is for the S2 equal to 2.6\%, which corresponds to $1.94\sigma$ (using Gaussian one-sided significance~\cite{Ganguly}) and is therefore only marginally significant. 

Therefore, even with this model for the errors, we do not see any uniformity in the location of the peak frequencies across the three samples. \textcolor{black}{However, in the S2 sample we do see a peak at 11/year similar to S16 and KN15, but with a lower significance than S16.}


\vspace{3mm}
\begin{table*}
\begin{tabular}{|c|c|c|c|}

 \hline
 Sample&Error model& Peak Frequency   & FAP  \\
 \hline
 S2& 0.03\%    &11.40 $yr^{-1}$&17.2\% \\
 S3&0.03\% &17.57 $yr^{-1}$&28.2\% \\
 S4&0.03\% &17.36 $yr^{-1}$&20.5\% \\
 S2&Std. error of Mean &11.22 $yr^{-1}$&2.6\% \\
 S3&Std. error of Mean &11.03 $yr^{-1}$&41.4\%\\
 S4&Std. error of Mean &1.33 $yr^{-1}$&74.1\%\\

 \hline
\end{tabular}
\caption{A summary of the results from the generalized LS analysis carried out on S2, S3, and S4 datasets using two different models for the errors per data point. The last two columns indicate the position of the peak frequency and  FAP. \textcolor{black} {We note that only S2 shows a marginally significant peak close to 11  per year when standard error of mean is used as the error model.}}
\label{deltabic}
\end{table*}

\section{Comparison with Sturrock et al}
\label{sec:scomp}
\textcolor{black}{In this section, we check if we can reproduce the results in Section 2 of  S16, where they dispute the significance calculation of KN16. For this purpose, we only focus on the data from the S2 sample, since this sample has the largest LS power at 11/year. We used the same binning procedure as our earlier analysis. Since the exact error model or the binning used to obtain LS power of 8.42  is not specified, we used both the error models. To compare our results with theirs, we use the same normalization for the LS power as in KN15 and S16 (which follows Scargle's convention~\cite{Scargle}), by multiplying the  power shown in Figs.~2-6 by $(N-1)$/2. We also calculate significance in the same way as Sect.2 of S16~\cite{NR}, and is given by $\exp(-S)$, where $S$ is the LS power using this normalization. 
This significance quantifies the false alarm probability of the null hypothesis
and is equivalent to a $p$-value.
In addition  to the generalized LS periodogram, we also calculate the normal LS power and its significance, to mimic the results of S16 as closely as possible.}

\textcolor{black}{Our results are shown in Table~\ref{compsturrock}. By positing an error model of 0.03\% per data point, we get a value for our significance about 100 times larger than that obtained in S16. The results don't differ much between normal and LS periodogram.
However, using standard error of the mean, we get a significance value about one order of magnitude smaller than that in S16 of about $1.6 \times 10^{-5}$. Taken at face value, this would correspond to 4.1$\sigma$ significance.
Therefore,  the actual value of the significance is also  sensitive to the choice of the error model used. Since the actual error model used in Section 2 of S16 is not explicitly specified,  we cannot do a direct comparison of our significance with theirs.}

\textcolor{black}{However, we can see that the statistical significance of the peak at 11/year becomes  enhanced compared to S16, using the second error model.}

\begin{table*}
\begin{tabular}{|c|c|c|c|c|}

 \hline
Sample & Error model& Periodogram  & LS Power (Scargle normalization) & Significance  \\
 \hline
 S2& 0.03\%    &Generalized&5.57&0.0038\\
 S2&0.03\% &Normal &5.52&0.0040\\
 S2&Std. error of Mean & Generalized &11.04&$1.6$x$10^{-5}$ \\
 S2&Std. error of Mean &Normal &11.04&$1.62$x$10^{-5}$\\

 \hline
\end{tabular}
\caption{Significance values of the peak computed at about $11 \rm{yr^{-1}}$ for the  S2 data sample using a generalized as well as normal LS periodogram. For comparison, S16~\cite{Sturrock16} (cf. Section 2 therein) finds a  value of LS power of 8.42 with significance of $2 \times 10^{-4}$.}
\label{compsturrock}
\end{table*}

\section{Is the data completely  stochastic?}
\label{sec:stochastic}
\textcolor{black}{Although our FAP is higher than S16, from Figs.~\ref{fig:S2e1} and ~\ref{fig:S2e2},  we do find a peak visible to the naked eye in the S2 data sample for both the choice of error models at the same frequency as S16. This raises the question of whether the observed data are purely stochastic.}

\textcolor{black}{Therefore, to test if the data are consistent with pure noise without any sinusoidal modulations, we carried out numerical experiments with synthetic data, using both the error models. We replaced the activity data of the sample S2 with Gaussian distributed  random numbers (which are proxy for the activity counts) at the same time instances when S2 had data, and carried out the power spectrum analysis and calculation of FAP in the same way as for real data. We generated random numbers with mean of zero and standard deviation of unity. We then replicated the above procedure of generating synthetic data and analyzing using LS periodogram 1000 times and constructed a histogram of  the LS power from each such realization. The LS power for each iteration was chosen as the maximum LS power in the frequency range between $11/yr$ and $11.5/yr$. Figures 8 and 9 depict the histograms of the LS power with $0.03\%$ error per data point  and with standard error of the mean respectively. We also  note that our results do not change much,  if we use the standard deviation of the original data. \\
We found that the random LS power rarely crosses the observed LS power. After 1000 trials, we find  that this maximum LS power value exceeds the observed LS power  at 11/year, for about 8  and 10 different realizations, for the 0.03\% error model and the standard error of mean error models respectively. Therefore, these numerical experiments with synthetic noise data demonstrate that the observed data are not completely stochastic and the observed LS power in the S2 data sample are indicative of  marginal hints for periodicity of 11 years.}
\begin{center}
\begin{figure*}
\includegraphics[scale=0.5]{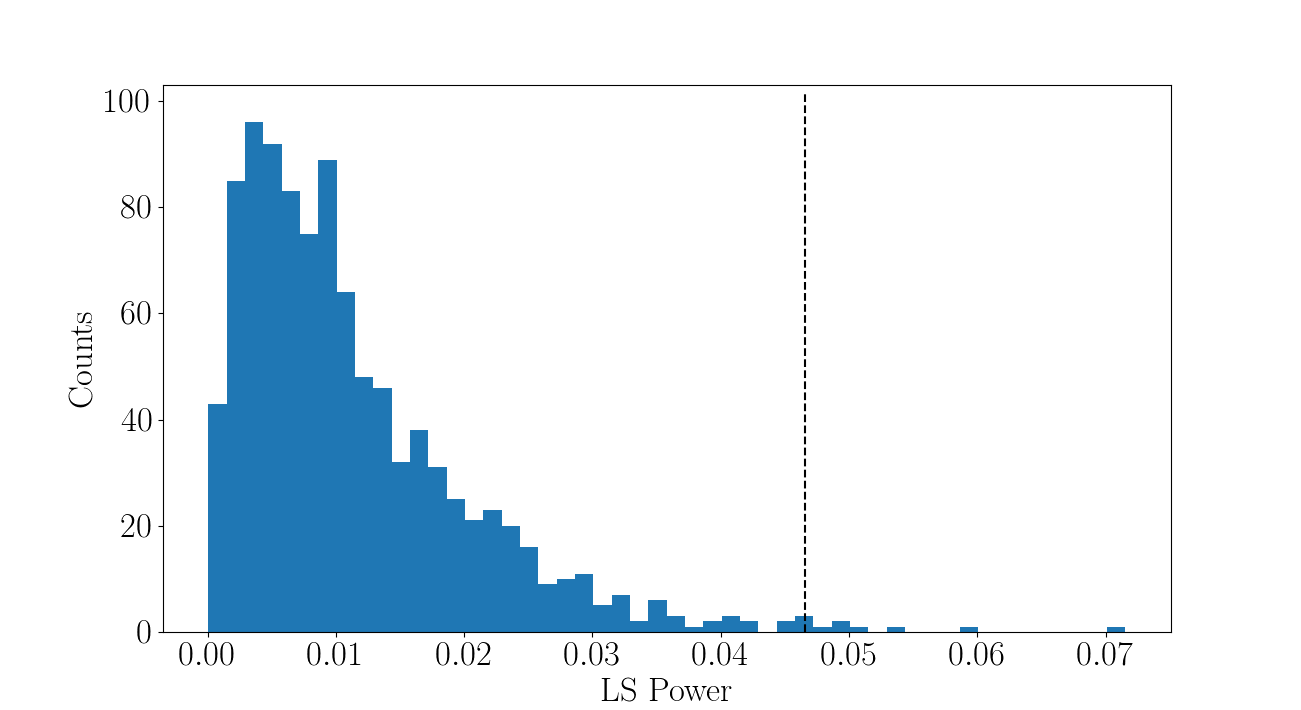}
\caption{Histogram representing the distribution of LS powers over 1000 iterations with random data. The black vertical dotted line represents the observed LS power with the data of sample S2. The above plot represents the analysis with $0.03\%$ error per data point as the error model. The probability of getting a peak larger than the observed value from these simulations is about 0.8\%.}
\label{fig8}
\end{figure*}
\end{center}
\begin{center}
\begin{figure*}
\includegraphics[scale=0.5]{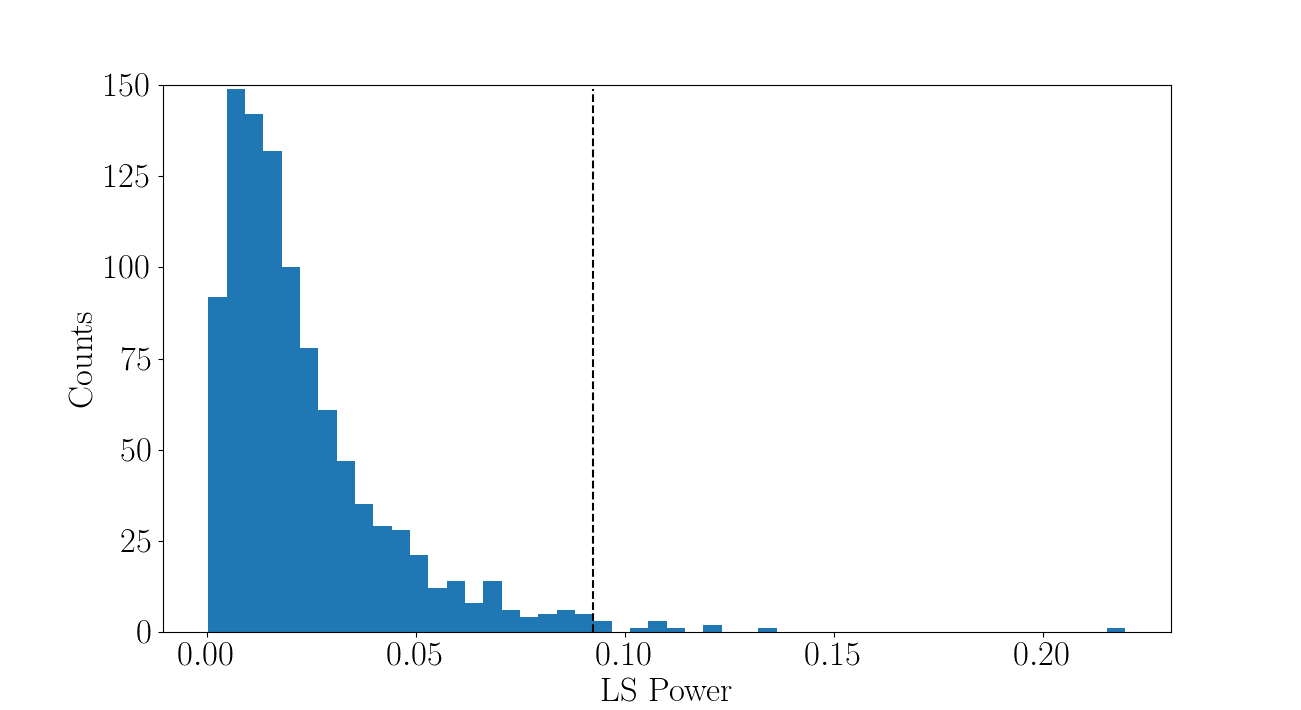}
\caption{Histogram representing the distribution of LS powers over 1000 iterations with random data. The black vertical dotted line represents the observed LS power with the data of sample S2. The above plot represents the analysis with Standard error of the mean as the error model. The probability of getting a peak larger than the observed value from these simulations is about 1\%.}
\label{fig9}
\end{figure*}
\end{center}

\section{Conclusions}
\label{sec:conclusions}
The aim of this work was  to resolve the controversy between two groups (S16 and KN15) regarding the influence of solar processes  on nuclear beta decay rates of $\rm{^{90} Sr/^{90}Y}$ measured at the  PTB.  We would like  to verify using these measurements, whether this decay mode shows sinusoidal variations with a frequency of 11/year as claimed by S16 (but disputed by KN15), which could be indicative of a solar influence. 

For this purpose, we have used the  generalized or floating-mean  LS  periodogram to search for periodicity in the PTB activity data 
for three different samples, for which measurements span a period of 400 days. This generalized LS periodogram has
been shown to be more sensitive than the normal periodogram, in case the data do not encompass the full phase coverage of a putative periodic signal~\cite{Vanderplas}.
We grouped the activity data into 240 bins, with each bin containing contiguous activity data points. We obtained the periodograms using two different assumptions about the errors as follows:
\begin{itemize}
\item 0.03\% error per data point (in accord with the error budget calculated in KN15).
\item Standard error of mean in each bin (similar to the analysis done in S16).
\end{itemize}
The significance of each peak was evaluated using bootstrap resampling with 1000 samples, using the method proposed by Suveges~\cite{Suveges}. 
The generalized LS periodograms for all the three samples are shown in Figures~2-7.
Table~\ref{deltabic} summarizes the results of the 
generalized LS analysis carried out on S2, S3, and S4 using the above mentioned error models.\\
\textcolor{black}{To compare our results with Sturrock et al, we then estimated the significance of the peak in the S2 sample using the same method as S16 with both the error models. Our results from this exercise are shown in Table~\ref{compsturrock}. We then addressed the question of whether the data are purely stochastic by conducting 1000 numerical experiments of activity time series, which are drawn from a normal distribution using the same  time-binning as the observed data. Histograms of the LS power at frequencies close to 11/year can be  found in Figs.~\ref{fig8} and \ref{fig9}.} 

Our conclusions about these analyses are as follows:
\begin{itemize}
\item  The peak frequencies and their significances  slightly change for some of the samples with different error models. 

\item We do not  find a peak in the periodograms  close to 11/year in all  the three samples using either of the two  error models. 

\item  The sample S2 has a peak at about 11 $yr^{-1}$ with FAP values of 17.2\% and 2.6\% assuming 0.03\% error per data point and standard error of mean respectively. The FAP of 2.6\% corresponds to a significance of 1.94$\sigma$, and \textcolor{black}{its statistical significance is smaller than that claimed in S16.}

\item The only other sample with a peak frequency close to 11/year   is S3, assuming a standard error of the mean. However, its  FAP of 41.4\%  is consistent with  a noise fluctuation.

\item None of the remaining peaks found in the other samples have FAP less than 10\% with either of the two error budgets. Therefore, none of them can be considered as evidence for sinusoidal variations in the beta decay rates.

\item \textcolor{black}{We obtain a significance of 0.4\% and 0.0016\%, using the same formula used by S16 for the 0.03\% error per point and standard error of the mean models respectively. These values  are about ten times larger and smaller respectively  than  the  significance of 0.02\% estimated in Section 2 of S16.}

\item \textcolor{black}{For purely stochastic time-series, we would obtain the probability of getting the LS power greater than the one observed at 11/year to be  about 1\%.}

\end{itemize}

Hence in conclusion, we see that the differently prepared chemical samples S2, S3, and S4 do not \textcolor{black}{exhibit any consistent periodic oscillations in their activity}. \textcolor{black}{However, we do see a marginally significant peak in the S2 data sample at the same frequency as S16 (11 per year), but with a higher false alarm probability. More data is needed for the S2 sample, along with a detailed error budget to ascertain if this peak at 11/year persists and is significant.}

\acknowledgement{
We are grateful to Karsten Kossert for providing us 
the data for the PTB measurements analyzed  in KN15 and answering our queries, which enabled us to do this analysis. All the plots in this work have been made using the {\tt astroML} library~\cite{astroml,proc}. We also thank the anonymous referee for detailed critical feedback on our manuscript. 
We also acknowledge Jake Vanderplas for useful correspondence.}

\bibliographystyle{spphys}
\bibliography{ref}

\end{document}